\documentclass[12pt]{article}
\usepackage{graphicx}
\newcommand{\bb}{\begin{equation}}
\newcommand{\ee}{\end{equation}}
\newcommand{\ba}{\begin{eqnarray}}
\newcommand{\ea}{\end{eqnarray}}

\begin{document}

\title{{\bf Critical Gravitational Inspiral\\ of Two Massless Particles}}

\author{
Don N. Page
\thanks{Internet address:
profdonpage@gmail.com}
\\
Department of Physics\\
4-183 CCIS\\
University of Alberta\\
Edmonton, Alberta T6G 2E1\\
Canada
}

\date{2022 October 5}

\maketitle
\large

\begin{abstract}
\baselineskip 18 pt

If two ultrarelativistic nonrotating black holes of masses $m_1$ and $m_2$ approach each other with fixed center-of-momentum (COM) total energy $E = \sqrt{s} \gg (m_1+m_2)c^2$ that has a corresponding Schwarzschild radius $R = 2GE/c^4$ much larger than the Schwarzschild radii of the individual black holes, here it is conjectured that at the critical impact parameter $b_c$ between scattering and coalescing into a single black hole, there will be an inspiral of many orbital rotations for $m_1c^2/E \ll 1$ and $m_2c^2/E \ll 1$ before a final black hole forms, during which all of the initial kinetic energy will be radiated away in gravitational waves by the time the individual black holes coalesce and settle down to a stationary state.  In the massless limit $m_1 = m_2 = 0$, in which the black holes are replaced by classical massless point particles, it is conjectured that for the critical impact parameter, all of the total energy will be radiated away by the time the two particle worldlines merge and end.  One might also conjecture that in the limit of starting with the massless particles having infinite energy in the infinite past with the correct ratio of impact parameter to energy, the spacetime for retarded time before the final worldline merger at zero energy will have a homothetic vector field and hence be self similar.  Evidence against these conjectures is also discussed, and if it proves correct, I conjecture that two massless particles can form any number of black holes. 

\end{abstract}

\normalsize

\baselineskip 22 pt

\newpage

\section{Introduction}

LIGO has had enormous success in detecting the gravitational waves from astrophysical inspiraling black holes that coalesce to form a larger black hole \cite{LIGOScientific:2016aoc,LIGOScientific:2018mvr,LIGOScientific:2020ibl,LIGOScientific:2021djp}.  
In these cases the black holes have been inspiraling long before their gravitational wave emission is strong enough to be detected by LIGO, so that well before the final coalescence, the two black holes are moving with nonrelativistic velocities $v \ll c$ relative to each other and are gravitationally bound to each other.

Astrophysically, it seems that it would be very rare (and not yet observed) for two black holes to approach at relativistic velocities $v \sim c$ before coalescing.  However, it is an interesting question what would happen if this occurred.  Indeed, there has been an enormous amount of work [5--67], especially recently, calculating, among other related things, approximations for the deflection angles $\theta$ and the gravitational radiation emitted by small objects (often approximated by point particles, though here for concreteness I shall assume they are nonrotating black holes if they have positive rest mass, or massless particles if not) approaching each other at arbitrary relative velocities in the limit of a sufficiently large impact parameter $b$ that the deflection angle $\theta$ is much less than one radian.

Here I wish to raise the question of what would happen at the critical impact parameter, $b = b_c$, depending on the black hole masses $m_1$ and $m_2$ (with total rest mass $M \equiv m_1+m_2$) and on the center-of-momentum (COM) total energy $E$, such that for $b > b_c$ the two black holes would scatter without coalescing, but for $b < b_c$ the two black holes would coalesce into a larger black hole.  I shall consider the ultrarelativistic case $M/E \ll 1$ and then focus on the limit $M/E = 0$, when the objects are massless particles instead of black holes.

I shall often use units in which the speed of light is set equal to unity, $c = 1$, but I shall write factors of Newton's constant $G$ explicitly and so not set it to unity.  Since I shall almost always be using classical physics, I shall not set $\hbar = 1$, though I shall take the classical limit 
$E \gg E_{\mathrm {Planck}} \equiv \sqrt{\hbar c^5/G} \approx 1.9561$ GJ,
and for black holes instead of massless particles, I shall assume
$m_1 \gg E_{\mathrm {Planck}}$ and $m_2 \gg E_{\mathrm {Planck}}$.

\section{Ultrarelativistic Collisions of Black Holes}

For $M \equiv m_1 + m_2 \ll E/c^2$ with initially Schwarzschild (nonrotating) black holes having positive rest masses $m_1$ and $m_2$, the critical impact parameter should be 
\bb
b_c = b_c(E,m_1,m_2) = (GE)\beta(m_1/E,m_2/E),
\label{beta}
\ee
with the dimensionless function $\beta(m_1/E,m_2/E)$, a symmetric function of its two dimensionless arguments, going to a constant, which I shall call $\beta_1$, in the limit that $m_1/E \rightarrow 0$ and $m_2/E \rightarrow 0$ (which is implied by the single limit $M/E \rightarrow 0$):
\bb
\beta_1 \equiv \lim_{M/E \rightarrow 0}\frac{b_c}{GE}.
\label{beta_1}
\ee 

For impact parameter $b$ just infinitesimally larger than the critical impact parameter $b_c$, the two black holes (each much smaller than the impact parameter, since their Schwarzschild radii, $2Gm_1$ and $2Gm_2$, are much smaller than $b_c \sim GE$, leaving out the dimensionless factor $\beta(M_1/E,m_2/E)$ that is expected to be of the order of unity) do not merge into a larger black hole, but I would expect them to radiate gravitational waves carrying off nearly all of the kinetic energy of the initial ultraralativistic black holes, though see below for alternative possibilities.

For impact parameter $b$ just infinitesimally smaller than the critical impact parameter $b_c$, the two black holes would merge into a larger black hole, and in this case I would expect all of the initial kinetic energy (and some of the gravitational potential energy released after the black holes have become bound) to be radiated as gravitational waves, unless the two black holes can grow greatly before merging.  In the limit that $b$ approaches $b_c$ from below and that one can neglect $m_1/E$ and $m_2/E$, I would expect that the final black hole mass would be 
\bb
M_f = M \mu(\nu)
\label{mu}
\ee
for some $O(1)$ dimensionless function $\mu(\nu)$ of the symmetric mass ratio function
\bb
\nu \equiv \frac{m_1m_2}{(m_1+m_2)^2} \equiv \frac{m_1m_2}{M^2} \leq \frac{1}{4}.
\label{nu}
\ee
Indeed, the Hawking area theorem \cite{Hawking:1971tu} for the initially nonrotating Schwarzschild black holes of mass $m_1$ and $m_2$ implies that $M_f^2 \geq m_1^2+m_2^2$ and hence that $\mu(\nu) \geq \sqrt{1-2\nu} \geq \sqrt{1/2}$.

If not too much of the emitted gravitational radiation enters the individual black holes, I would expect the radiation not only to carry away the initial kinetic energy but also some of the binding energy released when the two black holes become gravitationally bound before coalescing, and the coalescence itself would also release some additional energy into gravitational waves, so I would expect $\mu(\nu) < 1$ when $b_c - b \ll b_c$, so that the final black hole mass $M_f$ would be less than the sum of the initial black hole masses (as it is for nonrelativistic mergers such as those detected by LIGO), even when the initial total energy $E$ is much larger.  On the other hand, if the impact parameter $b$ is significantly less than $b_c$ (though perhaps still having $b_c-b \ll b_c$), I would expect that the final black hole mass would be, on a logarithmic scale, much closer to the total initial energy $E$ than to the total initial rest mass $M = m_1 + m_2 \ll E$ of the individual black holes.

It would be very interesting to try to find a good estimate for the dimensionless function $\mu(\nu) = M_f/M$ that is the ratio of the final black hole mass $M_f$ to the sum $M$ of the initial black hole masses in the limit that $M/E \rightarrow 0$, and to confirm that this limit exists.  (A finite limit for $\mu(\nu)$ would not exist if the two black holes absorb a significant amount of the initial kinetic energy that goes into gravitational radiation early on, a possibility that shall be discussed in more detail later.)  Assuming that a finite limit for $\mu(\nu)$ does exist, it would also be interesting to find low-order corrections to this ratio $\mu(\nu) = M_f/M$ when $m_1/E$ and $m_2/E$ are not infinitesimal.  However, both of these projects are far beyond the scope of this work.

\section{Collisions of High-Energy Massless Particles}

Next, let us consider the limit that $M = m_1 + m_2 \rightarrow 0$, but keeping the COM total energy $E \gg E_{\mathrm {Planck}}$, so that the collision can be treated as a classical collision with the gravitational interaction dominating.  (Indeed, I shall ignore all other interactions and just consider classical general relativity.)  I shall also assume that there are no incoming gravitational waves or other matter, so that the situation is uniquely determined by the initial positions and momenta of the two massless particles.

In this case the only relevant parameters are the COM total energy $E$ and the impact parameter $b = 2J/E$, where $J$ is the total angular momentum in the COM.

The factor of 2 occurs because for massless particles with trajectories in the COM frame that would have a minimum projected separation of $b$ if the particles moved in straight lines with no deflection, $b$ is twice the minimum projected distance of each particle from the COM, and each particle has an equal magnitude of spatial momentum (though the two momenta are in opposite directions and hence give total spatial momentum zero in the COM frame), $p = E/2$, in the COM, so the total orbital angular momentum in the COM is the sum of the two angular momenta of the two particles and hence is $J = 2(b/2)(E/2) = bE/2$.  Note that many papers on ultrarelativistic collisions take $E$ to be the energy of each particle in the COM, which is half my $E = \sqrt{s}$ with Mandelstam invariant $s = (p_1+p_2)\cdot(p_1+p_2) > 0$ (using metric signature $(+---)$ for simplicity in this paragraph only, but not later) for initial particle 4-momenta $p_1$ and $p_2$.  When $b = b_c = (GE)\beta_1$, $J = (1/2)GE^2\beta_1$.

In comparison with this orbital angular momentum $J \sim GE^2$, the spin angular momentum is some small integer, e.g., 1 or 2 for a photon or graviton respectively, multiplying $\hbar$, and by the assumption that $E \gg E_{\mathrm {Planck}} \equiv \sqrt{\hbar c^5/G}$, one gets $\hbar \ll GE^2/c^5 \sim J$, here including the factors of $c$ that I usually set equal to 1.  

Although there are therefore only the two relevant parameters, say $E$ and $b$, the Einstein equation of classical general relativity is invariant under rescaling (so long as energies are rescaled by the same factor as lengths in the assumed 4-dimensional spacetime, so that the stress-energy tensor components, which scale proportionally to the energy density, scale as the inverse squares of the lengths, as does the Einstein tensor on the other side of the Einstein equation).  Therefore, the classical situation in general relativity remains essentially invariant if both the COM total energy $E$ and the impact parameter $b$ are scaled by the same factor, so the classical physics depends nontrivially only on the dimensionless parameter $\beta \equiv b/(GE)$.

In the situation {\it not} being considered here of the collision of a photon of COM energy $E_1$ and a nonrotating black hole of rest mass $M \gg E_1$, so that in the COM frame the black hole, which has spatial momentum opposite the spatial momentum of magnitude $E_1$ of the photon, has COM energy $E_2 = \sqrt{M^2 + E_1^2} \approx M + (1/2) E_1^2/M$, then the total COM energy is 
\bb
E = E_1 + E_2 = E_1 + \sqrt{M^2 + E_1^2} = M + E_1 + (1/2) E_1^2/M + O(E_1^4/M^3) \approx M.
\label{EBH}
\ee
In this case, to lowest order in $E_1/M$, the critical impact parameter is \cite{Hagihara,Darwin}
\bb
b_c \approx \sqrt{27}GE \equiv (GE)\beta_0,
\label{bcSch}
\ee
with $\beta_0 = \sqrt{27}$ being the analogue of $\beta_1 \equiv b_c/(GE)$ (which is for both objects ultrarelativistic in the COM frame) when instead for $\beta_0$, one object (the black hole) is highly nonrelativistic ($v = E_1/\sqrt{M^2+E_1^2} \ll c = 1$) in the COM frame.  Nevertheless, just as I shall consider $\beta_0$ to be of the order of unity, so for $b_c$ being the critical impact parameter for the collision of two massless particles of total energy $E$ in the COM frame, I would expect the dimensionless constant $\beta_1 \equiv b_c/(GE)$ also to be of the order of unity, but the precise value is left to the reader to calculate if possible.

Analogous to Choptuik scaling \cite{Choptuik:1992jv}, one might conjecture \cite{Pretorius2022} that for impact parameter $b = (GE)\beta$ just slightly smaller than the critical impact parameter $b_c = (GE)\beta_1$, one has the ratio of the final black hole mass $M_f$ to the initial total COM energy $E$ obeying the relationship
\bb
M_f/E \propto (\beta_1 - \beta)^{\gamma_1}
\label{scaling}
\ee
for some critical exponent $\gamma_1$, which would be another dimensionless characteristic constant besides $\beta_1 \equiv b_c/(GE)$.

At the critical impact parameter $b = b_c = (GE)\beta_1$, I conjecture that the two classical massless point particles would spiral around each other an infinite number of times, while radiating away all their energy.  As the number of rotations increases indefinitely, I would expect that the motion would asymptotically become self-similar, approaching a logarithmic or equiangular spiral.

In flat space, a logarithmic spiral in polar coordinates $(r,\phi)$ has, with a constant $L$ of the dimensions of length and dimensionless polar slope $k = \tan{\alpha}$ in terms of the dimensionless polar slope angle $\alpha$, 
\bb
r = L e^{-k\phi} = L e^{-(\tan{\alpha})\phi}.
\label{spiral}
\ee
Here I have chosen a minus sign in the exponent so that if the polar angle $\phi$ is increasing, the radius $r$ is decreasing, so that the curve spirals in to the center.

However, the gravitational field of the inspiraling massless particles is expected to lead to significant curvature of the spacetime in the vicinity of the particles, so one needs to define what the constant-$r$ `spheres' are for giving the polar slope angle $\alpha$ as the deviation of the direction of the massless particle 3-momentum from being tangent to the `sphere' of constant $r$ at the same location.  One way to do this would be to choose the spatial origin, say at $r=0$, to be at the COM of the system (the center of momentum of the massless particles and the gravitational field), and let $t$ on the COM worldline be the proper time along the worldline, with the massless particle merger at $t=0$.  (Because of the symmetries of the situation, this worldline should be a timelike geodesic.  I shall assume that the singularity at the particle worldline merger, by which the particles have lost all their energy, is sufficiently weak that one can continue the spacetime beyond it, including the geodesic worldline of the COM for $t>0$.)  Then define the hypersurface for each value of $t$ by starting at that value of $t$ at the COM worldline and constructing spatial geodesics in all spatial directions out from that worldline orthogonal to it to form the hypersurface of constant $t$.  One can also define a radial coordinate $r_1$ to be the distance along each geodesic from the COM worldline at the spatial origin at $r_1=0$ and thereby define constant-$r_1$ `spheres' on the hypersurfaces of constant $t$.  

One can then use the constant-$t$ hypersurfaces and constant-$r_1$ `spheres' within each hypersurface to give a first definition of the polar slope angle, $\alpha_1$, and polar slope, $k_1 = \tan{\alpha_1}$.  In particular, project the 4-momentum of one of the massless particles at some $t$ and $r_1$ onto the tangent plane of the constant-$t$ hypersurface there to give a spatial 3-momentum $\bf{p}$ and unit (since at the speed of light, $c=1$) 3-velocity $\bf{v} = \bf{p}/|\bf{p}|$.  Then define the polar slope angle $\alpha_1$ to be the angle between that spatial 3-momentum and the tangent plane to the constant-$r_1$ `sphere' at that point.  In particular, if $\bf{n}$ is the outward unit 3-vector in the constant-$t$ hypersurface that is normal to the constant-$r_1$ `sphere' at that point, $\bf{n\cdot v} = -\sin{\alpha_1}$, so
\bb
k_1 = \tan{\alpha_1} = \frac{-\bf{n\cdot v}}{\sqrt{1-(\bf{n\cdot v})^2}}.
\label{k1}
\ee

It is conceivable that sufficiently far from the COM worldline, the geodesics emanating orthogonally from it may cross to give caustics and events with more than one such geodesic passing through it, and hence multiple $(t,r_1)$ coordinates for the same event.  I shall assume without proof that this does not happen for events along the massless particle worldlines, so that for each event along each of these worldlines, there is a unique set of $(t,r_1)$ coordinates.  However, even if this assumption is wrong and there are multiple sets of $(t,r_1)$ coordinates for an event on a massless particle worldline, one can choose the set with the smallest $r_1$, and if there are multiple sets with the same $r_1$ that is smallest, one can choose the one with the smallest value of $t$, thus getting a unique set of $(t,r_1)$ coordinates for events on each massless particle worldline, even if there are multiple geodesics emanating orthogonally from the COM to each such event on a massless particle worldline with this same set of values of $t$ and $r_1$.

One might also wish to give a full set of spherical polar coordinates $(t,r_1,\theta,\phi)$ for events sufficiently near the COM worldline that there are no caustics along the spatial geodesics, orthogonal to the COM worldline, between the COM worldline and the events of interest (which I shall continue to assume include the events along the massless particle worldlines).  A first requirement is that the metric very near the COM worldline should have the standard form (now using signature $(-+++)$)
\bb
ds^2 \approx - dt^2 + dr^2 + r^2(d\theta^2 + \sin^2{\theta} d\phi^2)
\label{metric1}
\ee
with the $\theta$ and $\phi$ coordinates constant along the geodesics that at $r=0$ are orthogonal to the COM worldline there.  At the COM worldline, the direction for each fixed value of $\theta$ and $\phi$ is parallel propagated along the geodesic COM worldline.  Therefore, worldlines at infinitesimal $r$ and fixed $\theta$ and $\phi$ are locally nonrotating.

With this coordinate restriction, one still has the freedom to make a time-independent (constant) global rotation of the angular $(\theta,\phi)$ coordinates, but this can be fixed so that the massless particles are spiralling around counterclockwise ($\phi$ increasing with time $t$) in the equatorial `plane' $\theta = \pi/2)$, and with $\phi$ fixed (modulo shifting by $\pi$, which interchanges the two identical massless particles and thus is a symmetry of the spacetime) so that when one of the particles has $r = L$ (which can be an arbitrary constant of the dimension of length), that particle is at $\phi = 0$.

In this way one can uniquely construct (up to the choice of the arbitrary length scale $L$ that just affects the location of the zero of $\phi$) a spherical polar coordinate form of Fermi coordinates around the COM worldline of the two particles and their gravitational field, giving a full set of coordinates that I shall now call $(t,r_1,\theta_1,\phi_1)$ in order to distinguish them from similar coordinates I shall define later.  (I shall not put a subscript on the $t$ coordinate, since in the later coordinate systems there is not a different $t$ coordinate; setting $t=0$ at the merger of the two massless particles, $t$ is always the proper time along the COM worldline at the event there from which a geodesic orthogonal to the COM worldline intersects the event off the COM worldline where $t$ is thus defined, assuming there is a unique such geodesic, or else using the paragraph after Eq.\ (\ref{k1}) above to pick unique values of $t$ and $r_1$.)  Note that by this method of construction, the $\phi_1$ coordinates are not rotating with respect to an inertial frame at the COM worldline, but by the dragging of inertial frames, they may be rotating with respect to static observers at infinity.

Now my conjecture is that in these Fermi coordinates and for the critical impact parameter $b= b_c = \beta_1GE$, as one approaches the final merger of the massless particles, the gravitational radiation emitted reduces their energy to zero, and they approach a self-similar behavior with $r_1(t) = -v_1t$, (where we recall that $t = 0$ is the coordinate time of the particle worldline merger), where 
\bb
v_1 = -\frac{dr_1}{dt}
\label{v1}
\ee
is a positive constant that can be interpreted to be an ingoing radial coordinate velocity of the two inspiraling massless particles.

In flat spacetime, $r = L e^{-k\phi}$ and $v = -dr/dt = krd\phi/dt$ give polar slope
\bb
k \equiv \tan{\alpha} = \frac{v}{rd\phi/dt},
\label{k}
\ee
so one can use this same equation in the $(t,r_1,\theta_1,\phi_1)$ coordinates to give a second definition of the polar slope, $k_2$, and polar slope angle, $\alpha_2$, in the curved spacetime of the two inspiraling massless particles, using the definition that $r_1(t)$ is the proper distance, from the spatial origin at the COM at $r_1=0$, to either of the two particles, along a geodesic of the spatial hypersurface of that value of $t$ (which is also a geodesic of the spacetime by the method of construction of the hypersurface as being generated by the spacetime geodesics orthogonal to the COM worldline).  That is, one may define
\bb
k_2 \equiv \tan{\alpha_2} = \frac{v_1}{r_1d\phi_1/dt} = -\frac{d\ln{r_1}}{d\phi_1}.
\label{k2}
\ee

This formula for the polar slope $k_2$ also has the consequence that if $r_{1a}$ and $r_{1b}$ are two successive values of $r_1(t)$ at the same $\phi_1$ for one of the massless particles, then
\bb
k_2 = \frac{1}{2\pi}\ln{\left(\frac{r_{1a}}{r_{1b}}\right)}.
\label{slope2b}
\ee

Therefore, in addition to the dimensionless critical impact parameter $\beta_1 = b_c/(GE)$, we now also have four other `independent' dimensionless constants, namely the critical scaling exponent $\gamma_1 = d\ln{M_f}/d\ln{(b_c-b)}$ at $b = b_c$, $v_1 = -dr_1/dt$, $k_1 = -\bf{n\cdot v}/\sqrt{1-(\bf{n\cdot v})^2}$, and $k_2 = -d(\ln{r_1})/d\phi_1$, presumably all of the order of unity, to be determined by numerical calculations that are far beyond the scope of this paper, but which I hope someone can be encouraged to do.  By `independent,' I mean they do not have obvious simple relations between them as $k = \tan{\alpha}$ and $\alpha$ do, but since all of them are determined by their definitions and by the unique solution of the Einstein equation near the zero-energy particle worldline merger point with initial conditions corresponding to two massless particles of total energy $E$ in asymptotically flat spacetime colliding with the critical impact parameter $b = b_c = (GE)\beta_1$ and with no incoming gravitational radiation, all of the dimensionless constants I have defined take logically determined values and so cannot be varied the way independent parameters usually can be; they just are what they are.

\section{Metric with a Homothetic Vector Field}

If one takes the conjectured asymptotically self-similar solution just before the merger and disappearance of the massless particles that by the time of their merger have radiated away all their energy, and extrapolates this solution back in time to remain self-similar all the way back to negative infinite time, one would have a solution that is self-similar for the entire spacetime before the future light cone of the particle worldline merger event.  This would effectively be the limit in which the massless particles have infinite impact parameter and infinite energy in the infinite past in the right ratio $G\beta_1$ so that there is an infinite period of self-similar inspiraling to the final particle worldline merger with then zero energy remaining in the particles.  All of the energy would be radiated outward in gravitational waves concentrated along the outward null geodesics from the worldline of the center of momentum (COM) that is at $r_1=0$.  As was assumed for the situation with two massless particles with finite total COM energy $E$ and finite impact parameter $b$, here for the precisely self-similar situation with infinite $b$ and infinite $GE$ in the fixed ratio $\beta_1 = b/(GE)$, I shall also assume no incoming gravitational radiation.

Let each of these outward null geodesics leaving the COM at $r_1=0$ at time $t < 0$ (at positive proper time $-t$ before the final particle worldline merger) be labeled by a dimensional retarded time null coordinate $U = t$ (equality only along the COM worldline at $r_1=0$) and by a dimensionless retarded time null coordinate $u = -\ln{(-U/L)}$ ($u = -\ln{(-t/L)}$ along the COM worldline, with an arbitrary constant length $L$ to set the scale for the dimensional quantities) that runs from $-\infty$ for null geodesics leaving $r_1=0$ at infinite past time, $t = -\infty$, to $+\infty$ for null geodesics leaving the particle worldline merger at $r_1=0$ and $t = 0$.  That is, at $r_1=0$, $t = U = -Le^{-u}$, increasing monotonically from $t = -\infty$ for $U = -\infty$ and $u = -\infty$ to $t = 0$ for $U = 0$ and $u = +\infty$.  Note that off the COM worldline, for $r_1 > 0$, one has $t > U = -Le^{-u}$, with $t$ increasing as one goes to greater $r_1$ along a null geodesic of fixed null coordinate $U$ or $u$.

The part of the spacetime where the retarded time null coordinate $u$ is defined is everything to the causal past of the future light cone of the particle worldline merger event at $r_1=0$, $t=0$, that light cone being given by $u = +\infty$.  This region where $-\infty < u < +\infty$ is the region where the spacetime is conjectured to be self similar, with a homothetic vector field
\bb
X = \frac{\partial}{\partial u}
\label{X}
\ee
and metric
\bb
ds^2 = L^2 e^{-2u}\tilde{g}_{ab}(\rho,\theta,\phi) dx^a dx^b
\label{metric}
\ee
with dimensionless coordinates $x^0 = u$, $x^1 = \rho = r/(-U)$, $x^2 = \theta$, and $x^3 = \phi$, with $(r,\theta,\phi)$ similar to the $(r_1,\theta_1,\phi_1)$ chosen as above but having one different variant for the radial coordinate $r = (-U)\rho$ and two different variants for the angular coordinates $(\theta,\phi)$, to be given below, and with the timelike coordinate $t$ replaced by the dimensionless null coordinate $u$ with $u$, $\theta$, and $\phi$ constant along outgoing null geodesics emanating from the COM worldline at $r_1=0$.  

The nontrivial part of the metric, $\tilde{g}_{ab}$, which depends only on the coordinates $(\rho,\theta,\phi)$ and not on the null coordinate $u$, is restricted to have $\tilde{g}_{11} \equiv \tilde{g}_{\rho\rho} = 0$, so that the trajectories with $u$, $\theta$, and $\phi$ constant are indeed null.  For these null trajectories also to be geodesics with affine parameter $\rho$, one can set $\tilde{g}_{01} \equiv \tilde{g}_{u\rho} = 1$ (which fixes the scale of $r = -U\rho$, so henceforth I shall call this dimensional radial coordinate $r_2 \equiv -U\rho \equiv Le^{-u}\rho$) and $\tilde{g}_{12} \equiv \tilde{g}_{\rho\theta} = \tilde{g}_{13} \equiv \tilde{g}_{\rho\phi} = 0$.  Then one can write the self-similar metric as
\ba
ds^2\!\!\!\!\! &=&\!\!\!\!\! L^2 e^{-2u}[-Adu^2-2dud\rho+2\rho du(Bd\theta+Cd\phi)
+\rho^2(Dd\theta^2+2Ed\theta d\phi+Fd\phi^2)] \nonumber \\
&=&\!\!\!\!\!-\!\left(\!A\!+\!\frac{2r_2}{-U}\!\right)dU^2\!-\!2dUdr_2\!
+\!2rdU(\!Bd\theta\!+\!Cd\phi)\!
+\!r^2(\!Dd\theta^2\!+\!2Ed\theta d\phi\!+\!Fd\phi^2)\!,
\label{metrica}
\ea
where $A$, $B$, $C$, $D$, $E$, and $F$ are six unknown functions of $\rho$, $\theta$, and $\phi$ (though not of $u$), making future numerical calculations easier than for a generic metric depending nontrivially on all four coordinates, but still highly nontrivial. 

One can choose (not uniquely) the $(\theta,\phi)$ angular coordinates so that at each fixed $u$ and $\rho$ they give a topological 2-sphere with metric proportional to
\bb
d\hat{s}^2 = Dd\theta^2+2Ed\theta d\phi+Fd\phi^2
\label{2-sphere}
\ee
that has the usual spherical polar coordinate ranges $0 \leq \theta \leq \pi$ and $0 \leq \phi < 2\pi$, with the limit as $\phi$ approaches $2\pi$ being the same point as $\phi = 0$ for each constant value of $\theta$, so that $\phi$ is periodic with period $2\pi$.  I shall also orient the north polar ($\theta = 0$), equatorial ($\theta = \pi/2$), and south polar ($\theta = \pi$) directions so that the massless particles, which are at constant $\rho = \rho_p$, are in the equatorial plane at $\theta = \pi/2$, and are at $\phi = 0$ (for what I shall call particle 1) and at $\phi = \pi$ (for what I shall call particle 2), with the particles rotating in the positive-$\phi$ direction in a nonrotating frame (counterclockwise as seen from the north pole).  However, note that the $\phi$ coordinate itself is rotating, relative to a nonrotating frame such as one at the COM worldline or one at future null infinity, $\mathcal{I}^+$, so although the two massless particles have fixed dimensionless coordinates $(\rho,\phi)$, namely $(\rho_p,0)$ for particle 1 and $(\rho_p,\pi)$ for particle 2, they are nevertheless rotating as seen from a nonrotating frame, and moving inward as measured by dimensionful coordinates such as the previously defined $r_1$ or the present $r_2 = Le^{-u}\rho$.

Because of the exchange symmetry between the two identical (except for location) massless particles, one can require that the six metric functions $A$, $B$, $C$, $D$, $E$, and $F$ in Eq.\ (\ref{metrica}) remain the same under the replacement of $\phi$ with $\phi+\pi$, which interchanges the coordinates of the two massless particles.
One can also require that the coordinates give manifest reflection symmetry about the equatorial plane, the same spacetime geometry when one interchanges $\theta$ with $\pi-\theta$, namely that at fixed $\rho$ and $\phi$ where the six metric functions in Eq.\ (\ref{metrica}) are just functions of $\theta$, they obey 
$A(\theta) = A(\pi-\theta)$, $B(\theta) = -B(\pi-\theta)$, $C(\theta) = C(\pi-\theta)$, $D(\theta) = D(\pi-\theta)$, $E(\theta) = -E(\pi-\theta)$, and $F(\theta) = F(\pi-\theta)$.

I shall assume as usual that the fixed points of $\partial/\partial\phi$ (which is {\it not} a Killing vector of this spacetime, since the massless particles at $\phi = 0$ and $\phi = \pi$, and their gravitational fields, break the rotational symmetry) are at the north pole, $\theta = 0$, and at the south pole, $\theta = \pi$.  At each fixed $u$ and $\rho$, I shall assume that as one approaches either pole, the metric function $D$ approaches a positive constant, that $E$ approaches 0, and that $F$ approaches zero.  To avoid conical singularities at the poles, I shall also require, as usual, that the circumferences of the circles of the dimensionless 2-sphere metric (\ref{2-sphere}) at fixed $u$,$\rho$, and $\theta$, namely $c(u,\rho,\theta) = \int_0^{2\pi}\sqrt{F(\rho,\theta,\phi)}d\phi$, approach $2\pi\sqrt{D}\theta$ at the north pole and $2\pi\sqrt{D}(\pi-\theta)$ at the south pole, so that the proper distance circumference $c(u,\rho,\theta)$ approaches $2\pi$ times the proper radial distance to the nearest pole as one approaches either of the two poles of $\theta$ at each constant value of $u$ and $\rho$ for the 2-dimensional metric (\ref{2-sphere}) and hence also for the full 4-dimensional metric (\ref{metrica}).

Geometrically, the new radial coordinate $r_2 = Le^{-u}\rho$, as defined above, is an affine parameter along the null geodesics of constant $u$, $\theta$, and $\phi$, with $r_2 = 0$ at the COM worldline, and increasing along the outward null geodesics with constant $u$, and normalized so that just as they leave the COM worldline, they have $dr_2/dr_1 = 1$.  Then $r_2 = L e^{-ku} \rho = r_1 + O(r_1^3)$, so the two $r$ coordinates are very nearly the same close to the COM worldline but might differ significantly along the massless particle worldlines and would probably differ greatly when either $r$ is much greater than $Le^{-u}$, which is the proper time remaining before the particle worldline merger on the COM at the same value of $u$ as the location where $r$ (either $r_1$ or $r_2$) is evaluated.

Assuming no incoming gravitational waves, the metric should be asymptotically flat at future null infinity, $\mathcal{I}^+$, where $r_2 = Le^{-u}\rho \rightarrow \infty$ at constant retarded time $u$, though it certainly is not asymptotically flat at spatial infinity, say as $r_1$ and $r_2$ are taken to infinity along one of the spatial geodesics at constant $t$ that are orthogonal to the COM worldline, since in that direction the mass interior to a `sphere' of constant $r_1$ or $r_2$ would grow proportionally to $r_1$ or $r_2$ and become infinite at spatial infinity.  On $\mathcal{I}^+$, the self similarity implies that the Bondi mass should have the form
\bb
M_B(u) = mLe^{-u}
\label{Bondi}
\ee
with a sixth dimensionless constant, $m$, to be determined by numerical calculations.

Just as the constant radial coordinate inward velocity $v_1 = -dr_1/dt$ along one of the massless particle worldlines was in terms of the radial coordinate $r_1$ (the orthogonal proper distance from the COM worldline) and the time $t$ which is constant along the spatial geodesics orthogonal to the COM worldline, so one can also define constant radial coordinate inward velocities of the radial coordinate $r_2$ that is a normalized affine parameter along the outgoing null geodesics of constant $U$ and $u$, differentiated with respect to the time $t$ or with respect to the retarded time $U$ that is constant along the outgoing null geodesics emanating from the COM worldline, with both the radial coordinate ($r_1$ or $r_2$) and the time coordinate ($t$ or $U$) taken along one of the massless particle worldlines:
\bb
v_2 = -\frac{dr_2}{dt},
\label{v2}
\ee
\bb
v_3 = -\frac{dr_1}{dU},
\label{v3}
\ee
\bb
v_4 = -\frac{dr_2}{dU}.
\label{v4}
\ee
The constant $v_4$ is also the constant value of the dimensionless radial coordinate $\rho = \rho_p$ at the worldline of each massless particle.

We can also get other constants analogous to the polar slope $k_2 = -d(\ln{r_1})/d\phi_1$, but to give a larger range of possibilities, let me first give two variant definitions of the angular coordinates $(\theta,\phi)$ in addition to the previous one that led to the metric of the form given by Eq.\ (\ref{metric1}) near the COM.  In particular, we can define locally nonrotating angular coordinates $(\theta_2,\phi_2)$ so that near the COM worldline one has
\ba
ds^2 &\approx&  - dT^2 + dr_2^2 + r_2^2(d\theta_2^2 + \sin^2{\theta_2} d\phi_2^2)
\nonumber \\
&=& - dU^2 - 2dUdr_2 + r_2^2(d\theta_2^2 + \sin^2{\theta_2} d\phi_2^2)
\label{metric2}
\ea
with $T \equiv U + r_2$, and then choose the orientation of $(\theta_2,\phi_2)$ so that the massless particles spiral inward counterclockwise ($\phi_2$ increasing) in the equatorial `plane' (which, in each 3-dimensional hypersurface of constant $t$ that is generated by the spatial geodesics emanating orthogonally from the COM worldline, is a 2-dimensional surface that has zero extrinsic curvature but that would be expected to have nonzero intrinsic curvature that presumably has both radial and angular dependence, on both $\rho$ and $\phi$, perhaps diverging at the location of each massless particle worldline) with $\theta_2 = \pi/2$, and $\phi_2 = 0$ where one of the particles is at $u=0$, and there has $r_2 = L\rho = L\rho_p$.  Like the $(\theta_1,\phi_1)$ angular coordinates, these will be nonrotating relative to an inertial frame at the COM worldline at $r_2 = 0$, but they almost certainly will be rotating relative to a frame at infinity.

Now an analogue of the polar slope $k_2 = -d(\ln{r_1})/d\phi_1$ for the inspiraling massless particles that remain at fixed $(\rho,\theta,\phi)$ is
\bb
k_3 \equiv -\frac{d(\ln{r_2})}{d\phi_2} = -\frac{d[\ln{(Le^{-u}\rho)}]}{d\phi_2}
= \frac{du}{d\phi_2}.
\label{k3}
\ee
This gives the relation between the rotating angular coordinate $\phi$ in the self-similar metric (\ref{metrica}) with the homothetic vector $X = \partial/\partial u$ and the coordinate $\phi_2$ that is locally nonrotating at the COM:
\bb
\phi_2 = \phi + u/k_3.
\label{phi2}
\ee 

An alternative choice of the angular coordinates, $(\theta_3,\phi_3)$, will have the metric approach the form
\ba
ds^2 &=& - dT^2 + dr_2^2 + r_2^2(d\theta_3^2 + \sin^2{\theta_3} d\phi_3^2)
\nonumber\\
&=& - dU^2 - 2dUdr_2 + r_2^2(d\theta_3^2 + \sin^2{\theta_3} d\phi_3^2)
\label{metric3}
\ea
at future null infinity, $\mathcal{I}^+$, where $r_2 \rightarrow \infty$ at constant retarded time $u$, similarly choosing the orientation of $(\theta_3,\phi_3)$ so that the massless particles spiral inward counterclockwise ($\phi_3$ increasing) in the equatorial `plane' with $\theta_3 = \pi/2$, and $\phi_3 = 0$ where one of the particles has $r_2 = L$.  These $(\theta_3,\phi_3)$ coordinates that are nonrotating at $\mathcal{I}^+$ will, almost certainly, because of the dragging of inertial frames, be rotating relative to the $(\theta_2,\phi_2)$ coordinates that give a nonrotating frame at the COM, and one may use them to give a fourth definition of the polar slope as
\bb
k_4 \equiv -\frac{d(\ln{r_2})}{d\phi_3} = -\frac{d[\ln{(Le^{-u}\rho)}]}{d\phi_3}
= \frac{du}{d\phi_3}.
\label{k4}
\ee
This will then give
\bb
\phi_3 = \phi + u/k_4.
\label{phi3}
\ee

Besides having $\phi_3$ rotating relative to $\phi_2$ as the result of the dragging of inertial frames from the angular momentum of the massless particles and the gravitational radiation they emit, the shear of the congruence of outgoing null geodesics at fixed $(u,\theta,\phi)$, from the Weyl curvature of the nonflat spacetime that is vacuum outside the massless particle worldlines, will cause $\theta_3$ to be distorted relative to $\theta_2$, though the interchange symmetry of the two massless particles will result in the fact that the polar directions are the same for both sets of coordinates ($\theta_2 = 0$ at $\theta_3 = 0$ and $\theta_2 = \pi$ at $\theta_2 = \pi$), and the equatorial planes are the same ($\theta_2 = \pi/2$ at $\theta_3 = \pi/2$).  However, although $4\pi\sin^2{(\theta_2/2)}$ is the solid angle inside the cone of interior angle $\theta_2$ near the COM, this is not the case at $\mathcal{I}^+$, where instead $4\pi\sin^2{(\theta_3/2)}$ is the solid angle inside the cone of interior angle $\theta_3$.  Alternatively, one may note that plotting $\theta_3$ versus $\theta_2$ will give a curve, symmetric under a rotation of $\pi$ radians about its midpoint at $\theta_2 = \theta_3 = \pi/2$, and whose integral, say $S$, of the square of the curvature of this curve over its length would be another dimensionless geometric invariant of the spacetime.

Other constants of the self-similar metric solution, with the two massless particles spiralling in from infinite energy and angular momentum at infinite separation in the infinite past, would come from hypothetical observations of a nonrotating observer at the north pole ($\theta_2 = \theta_3 = 0$) of $\mathcal{I}^+$.  From the viewpoint of such an observer that stays at constant $\theta_2 = \theta_3 = 0$ and at constant $r_2 \gg - U$, the angular separation of the inspiraling massless particles, as given by hypothetical infinitesimal-wavelength photons propagating along null geodesics from the massless particle worldlines to the observer worldline, will be very small, but multiplying this by $r_2 = -U\rho$ will give a value that at fixed null coordinate $U = -Le^{-u}$ at the observer tends, in the limit that the observer's $r_2$ is taken to infinity so that it is taken to $\mathcal{I}^+$, to a constant I shall call $2\Delta r(U)$.  Then $\Delta r(U)$ is the value for this method of calculating the separation of each massless particle from the COM, which may be called the `observed radius' of each particle as seen by the observer at $\mathcal{I}^+$ for each value of $U$.

In a nonrotating frame at the location of the observer, the observed direction of particle 1 from the COM in the plane perpendicular to the line of sight can be described by an angle I shall call $\varphi$.  Then one can make a fifth definition of the polar slope angle as
\bb
k_5 \equiv -\frac{d(\ln{\Delta r})}{d\varphi},
\label{k5}
\ee
which is the polar slope angle of the inspiral as observed by a nonrotating observer on the polar axis at future null infinity, $\mathcal{I}^+$.

One can also define a fifth constant `radial velocity'
\bb
v_5 = -\frac{d\Delta r}{dU},
\label{v5}
\ee
where $U$ is the dimensional retarded time $Le^{-u}$ at the observer when observing the 'observed radius' $\Delta r$ of each particle.

A sixth constant `radial velocity' may be defined to be
\bb
v_6 = -\frac{d\Delta r}{dU'},
\label{v6}
\ee
where $U'$ is the value of the dimensional retarded time at the observed massless particle when it emits the hypothetical photon that is observed by the observer at retarded time $U$.  Because the null geodesic of retarded time $U'=t'$ from the COM at its proper time $t'$ to the worldline of massless particle 1 is not in the same direction as the null geodesic of retarded time $U=t$ from the COM at its proper time $t$ to the observer at the time there when the observer sees the photon from the particle 1 worldline at retarded time $U$, $t'$ will be before $t$ and hence $U' < U$, but with a difference $U-U'$ that is essentially the time $\Delta t = t-t'$ it takes for the null geodesic of retarded time $U'$ to propagate from the COM at time $t'$ to particle 1 at time $t$.  Since this difference is decreasing as particle 1 spirals inward, $dU/dU' < 1$ and hence $v_6 < v_5$.

For a seventh definition of a constant `radial velocity,' instead of using the retarded time $U$ at the observer's reception of the hypothetical photon from the massless particle worldline, or the retarded time $U'$ of the massless particle worldline when it emits the hypothetical photon that is seen later by the observer, one can use the proper time $\tau$ measured by the observer (who is restricted to stay at fixed $r_2$ and fixed $\theta_3 = 0$, and hence be static in the asymptotically nonrotating coordinates $(U,r_2,\theta_3,\phi_3)$).  Then one can define
\bb
v_7 = -\frac{d\Delta r}{d\tau}.
\label{v7}
\ee
Because of time dilation near the COM relative to infinity, which presumably gives $dU/d\tau < 1$, I would expect that $v_7 < v_6 < v_5$.

It would be very interesting if one could find numerically the self-similar solution (with homothetic vector $X = \partial/\partial u$) of the Einstein equation to determine the six functions $A$, $B$, $C$, $D$, $E$, and $F$ of the three nontrivial coordinates (the dimensionless radial coordinate $\rho$ and two angular coordinates, say either $(\theta_2,\phi_2)$ or $(\theta_3,\phi_3)$) in the metric (\ref{metrica}), when the spacetime is vacuum except for two inspiraling massless particles at fixed $\rho = \rho_p$, $\theta_2 = \theta_3 = \pi/2$, and $\phi_2 = \phi_3 = 0$ or $\pi$.  With a suitable numerical analysis of this metric, one could evaluate a number of its dimensionless constant geometric invariants, such as the dimensionless critical impact parameter  $\beta_1 = b_c/(GE)$, the critical scaling exponent $\gamma_1$, the `polar slopes' $k_1$, $k_2$, $k_3$, $k_4$, and $k_5$, the `radial velocities' $v_1$, $v_2$, $v_3$, $v_4$, $v_5$, $v_6$, and $v_7$, the dimensionless rescaled Bondi mass $m$, and the integrated squared curvature $S$ of the $\theta_3(\theta_2)$ curve, which give 16 different gauge-fixed and hence diffeomorphism-invariant dimensionless positive real number characteristics of the self-similar metric with no incoming radiation or other matter but just two inspiraling massless particles that lose all their energy by the time of their final merger at $t=0$.

\section{Multiple Black-Hole Formation?}

After the first version of this paper was posted on the arXiv, Vitor Cardoso informed me of the evidence in \cite{Sperhake:2012me}, that for nonspinning black holes with total COM energy $E = 2.49M$ with here $M$ being the total of the rest masses of the holes (not to be confused with the $M$ in \cite{Sperhake:2012me}, which is the total energy of each black hole in the COM, half the total COM energy, or $E/2$ in my notation), only about 57.4\% of the initial kinetic energy is radiated, and about 32\% is absorbed by the individual black holes before they merge.  An extrapolation to infinite $E/M$ of this and other data with $E/M < 2.49$ gave about 41\% radiated, with an upper bound of about 70\%.  Although $E/M = 2.49$ hardly seems large enough for making firm conclusions about infinite $E/M$, the paper summarizes its findings by concluding, ``Our simulations thus settle the long-standing question of whether it is possible to release all of the c.m.\ energy as GWs in high-energy BH collisions: the answer is no.''

Huan Yang kindly informed me of Frans Pretorius's lecture at KITP 2022 May 20 \cite{Pretorius2022}, which seemed much more agnostic as to whether or not all of the kinetic energy would be radiated away in an asymptotically self-similar solution (which I now see he conjectured publicly before my paper).  For the collision of two massless particles, Pretorius explained that the process for avoiding radiating away all of the kinetic energy would be for the collision of the two pp-waves of the Aichelburg-Sexl metric \cite{Aichelburg:1970dh} for each initial massless particle to produce two initially separate black holes for impact parameter $b < (GE)\beta_2$.  Assuming that $\beta_2$ is greater than $\beta_1$, which now is taken to be the supremum of the values of $b/(GE)$ that lead eventually to one black hole, for $b < (GE)\beta_1 < (GE)\beta_2$, the two black holes would merge to form one hole, but for $(GE)\beta_1 < b < (GE)\beta_2$, the two black holes would not merge but would instead escape to infinity unbound to each other, and for $(GE)\beta_2 < b$, no black hole would form at all.

In this case, $\beta_1 < \beta_2$, the mass of the single black hole that would eventually form for impact parameter $b$ infinitesimally smaller than $(GE)\beta_1$ would not be infinitesimally small as it would for the inspiral of the two massless particles conjectured above when one does not have $\beta_1 < \beta_2$, but I would expect it to approach a nonzero fraction $f$ of the initial total COM energy, $M_f = fE$, as one takes the limit of the impact parameter $b$ approaching $(GE)\beta_1$.  (For $b = (GE)\beta_1$, I would expect the final state to be two black holes receding from each other at precisely their escape value, always slowing down but never coming to rest to fall back together to form a single black hole.)  Therefore, in this case one would not have the scaling relation $M_f/E \propto (\beta_1 - \beta)^{\gamma_1}$ of Eq.\ (\ref{scaling}).  On the other hand, for $b$ infinitesimally smaller than $(GE)\beta_2$, I might expect
\bb
M_f/E \propto (\beta_2 - \beta)^{\gamma_2}
\label{scaling2}
\ee
for a different critical exponent $\gamma_2$ that replaces $\gamma_1$ that would apply if it is not true that $\beta_1 < \beta_2$.

Having defined $\beta_1$ to be the supremum of values of $b/(GE)$ for two massless particles of initial total COM energy $E$ and impact parameter $b$, with no other matter or gravitational waves initially present, eventually to form a single black hole, and $\beta_2$ to be the supremum of values of $b/(GE)$ that form two black holes that escape unbound to each other, we may also contemplate the possibility of $n > 2$ black holes forming, with $\beta_n$ the supremum of values of $b/(GE)$ that form $n$ black holes that all escape unbound to all the other black holes.  This sounds even more difficult to achieve than to form just $n=2$ black holes unbound to each other, but if the two massless particles (or the black holes they form) can orbit around each other long enough and in the right configuration so that their mutual nonlinear gravitational wave field can gravitationally collapse into more black holes, I do not see that it would be obviously impossible.  If $n=2$ is possible, why not all $n$?

Now if indeed two classical massless point particles can form any number $n$ of unbound black holes, and if the critical values $\beta_n$, the supremum values of $b/(GE)$ that form $n$ black holes unbound to each other, obey the inequality $b_n < b_{n+1}$ for all positive integers $n$, then there would be an infinite set of ranges of impact parameters $b$, namely $(GE)\beta_n < b < (GE)\beta_{n+1}$ for each $n$, such that precisely $n+1$ unbound black holes are formed.  In this case one could also define $\beta_\infty$ to be the supremum of the $\beta_n$ values, so that for $b > (GE)\beta_\infty$ no black holes form at all.

If indeed there is this infinite discrete increasing sequence of critical dimensionless impact factors $\beta_n$ with a finite supremum $\beta_\infty$, one would like to learn their values and their asymptotic behavior.  A rather simple asymptotic behavior that one might conjecture is that
\bb
\beta_\infty - \beta_n \sim a x^n
\label{asymp}
\ee
for some constants $a$ and $x$, with $x < 1$ being the asymptotic ratio of the ranges of the impact parameters for $n+1$ unbound black holes and for $n$ black holes.

On the other hand, it still seems plausible to me that if $\beta_1$ is the supremum of $b/(GE)$ that can form one black hole, for $b > (GE)\beta_1$ no black holes can form at all, whether or not more than one unbound black hole can form for any $b < (GE)\beta_1$.  In this case, I would think that the conjectures in the previous sections still could be true, so that for nonspinning black holes with rest masses $m_1$ and $m_2$ and total rest mass $M\equiv m_1+m_2 \ll E$, where E is the total center-of-momentum energy, the two black holes could inspiral to form a single black hole of final mass $M_f = M\mu(\nu) < E$, and so that if one instead starts with two classical massless point particles of total COM energy $E$ at the critical impact parameter $b = b_c = (GE)\beta_1$, they inspiral to give an asymptotically self-similar metric, losing all their energy to gravitational radiation by the time their worldlines merge and end.

\section{Conclusions}

In this paper I have conjectured, first, that two nonrotating black holes of total rest mass $M \equiv m_1 + m_2 \ll E$, where $E$ is their total energy in the center-of-momentum (COM) frame, have a critical impact parameter $b_c = (GE)\beta(m_1/E,m_2/E)$, with $\beta$ a symmetric dimensionless function of its two dimensionless arguments, at which the two black holes spiral around each other until they lose to gravitational radiation all of their initial kinetic energy (and also some of the potential energy released when they become gravitationally bound), before forming a black hole of final mass 
$M_f = M \mu(m_1m_2/M^2)$ with some dimensionless function $\mu(\nu) \geq \sqrt{1-2\nu} \geq \sqrt{1/2}$ of its one dimensionless argument $\nu = m_1m_2/(m_1+m_2)^2 \leq 1/4$.

Second, I have conjectured that in the limit $M/E = 0$, so that the two black holes are replaced by classical massless point particles of COM total energy $E$, at the critical impact parameter $b_c = (GE)\beta_1$ for some dimensionless constant $\beta_1$, the two massless particles inspiral to an asymptotically self-similar configuration as they lose all their energy to gravitational radiation before colliding.  When one extrapolates the asymptotically self-similar solution infinitely back in time to get an exactly self-similar solution (with a homothetic vector field) that in the infinite past has the massless particles with infinite energy and infinite impact parameter in the right ratio, but no incoming gravitational radiation or any other matter, one gets a metric depending nontrivially on only three of the four coordinates, which can be characterized by a large number of dimensionless constants (geometric invariants defined in terms of the unique solution), of which I have defined 16 in this paper.

Third, in the case that more than one unbound black hole can form from the collision of massless particles, I have conjectured that any number $n$ can form.

\section*{Acknowledgments}

I am grateful for an email discussion with Gabriele Veneziano, in which, among other helpful information, he sent me the slides of his recent KITP lecture \cite{Veneziano:2022KITP}, provided many relevant references, and noted a confusion with my use of ``collision,'' which I have now often replaced with ``particle worldline merger.''  I thank Vitor Cardoso for informing me of the evidence in \cite{Sperhake:2012me} against my conjecture that all the kinetic energy gets radiated away, a crucial assumption for the conjectured asymptotically self-similar behavior.  Finally, Huan Yang kindly informed me of Frans Pretorius's May 2022 lecture at KITP, which I had not seen before and which predates some, though not all, of the conjectures in this paper.  This research was supported in part by the Natural Sciences and Engineering Research Council of Canada.


\end{document}